\documentclass[letterpaper, 10 pt, conference]{ieeeconf}  

\IEEEoverridecommandlockouts                              
\renewcommand{\O}{\mathcal{O}}
\newcommand{\vett}[1]{\left[\begin{matrix} #1 \end{matrix}\right]}
\newcommand{\Real}{\mathbb{R}}
\newtheorem{proposition}{Proposition}
\usepackage{algorithm}
\usepackage{algpseudocode}
\usepackage{graphics} 
\usepackage{epsfig} 
\usepackage{mathptmx} 
\usepackage{times} 
\usepackage{amsmath} 
\usepackage{amssymb}  
\usepackage{siunitx}
\usepackage{xcolor}
\usepackage[skip=1pt]{caption}

\renewcommand{\baselinestretch}{0.96}

\title{\LARGE \bf
A Branch and Bound method for the exact parameter identification of the PK/PD model for anesthetic drugs}

\author{Giulia Di Credico, Luca Consolini, Mattia Laurini, Marco Locatelli,\\
Marco Milanesi, Michele Schiavo and Antonio Visioli
\thanks{Luca Consolini, Giulia Di Credico, Mattia Laurini and Marco Locatelli are with Dipartimento di Ingegneria e Architettura, Universit\`a degli Studi di Parma, Italy.
{\tt\footnotesize \{giulia.dicredico, luca.consolini, mattia.laurini, marco.locatelli\}@unipr.it}}%
\thanks{Marco Milanesi, Michele Schiavo and Antonio Visioli are with Dipartimento di Ingegneria Meccanica e Industriale, Universit\`a degli Studi di Brescia, Italy. {\tt\footnotesize \{marco.milanesi, michele.schiavo, antonio.visioli\}@unibs.it}}
\thanks{This work has been developed under the PRIN 2022 project ``ACTIVA-Automatic Control of Total IntraVenous Anesthesia'', CUP D53D23001180006 funded by European Union – Next Generation EU.}
}

\begin{document}

\maketitle
\thispagestyle{empty}
\pagestyle{empty}

\begin{abstract}
We address the problem of parameter identification for the standard pharmacokinetic/pharmacodynamic (PK/PD) model for anesthetic drugs. Our main contribution is the development of a global optimization method that guarantees finding the parameters that minimize the one-step ahead prediction error. The method is based on a branch-and-bound algorithm, that can be applied to solve a more general class of nonlinear regression problems.
We present some simulation results, based on a dataset of twelve patients. In these simulations, we are always able to identify the exact parameters, despite the non-convexity of the overall identification problem.
\end{abstract}

\vspace{-15pt}
\section{Introduction}
Anesthesia provides a suitable level of depth of hypnosis (DoH), analgesia, and neuromuscular blockade (NMB) to patients.
In particular, in total intravenous anesthesia (TIVA), each of these effects is regulated by a specific drug.
The bispectral index (BIS) is widely employed to measure the DoH.
It is based on the analysis of the electroencephalogram (EEG), resulting in a dimensionless number between 0, corresponding to EEG silence, and 100, corresponding to a fully awake patient.
During surgical procedures, a target range between 40 and 60 is suggested to prevent awareness and to reduce the dose of anesthetic agent. An optimal depth of sedation is a main determinant of the quality of postoperative recovery.
Indeed, insufficient sedation exposes patients to awareness, with potential long-term psychological consequences, while excessively deep anesthesia induces hypotension, which is independently associated with increased postoperative morbidity and mortality.
In intensive care units (ICUs), excessive sedation in critically-ill patients, suffering from acute respiratory distress syndrome (ARDS), is associated with poor outcome and delirium.

Model-based control techniques, such as feedforward/feedback control, or model predictive control, leverage the knowledge of the pharmacokinetic/pharmacodynamic (PK/PD) model.
The PK/PD model describes the evolution in time of the effect of the hypnotic drug on the BIS signal. PK describes the dynamics of the drug concentration in the human body, while PD describes the relationship between the drug concentration and the clinical effect.
It has the structure of a Wiener model, composed of the cascade of a linear PK system and an algebraic nonlinear PD system~\cite{HODREA2012}.
The parameters of the linear part can be roughly estimated from the patient demographic data.
The parameters of the PD system, related to the patient's sensitivity to the hypnotic agent, are more difficult to estimate.

\subsection{Related literature}
There is a quite extensive literature on the identification of the PK/PD model of drugs used in general anesthesia.
Some works use linear regression to relate the parameters of the PK model to some of the patient's characteristics, such as age, sex, and body weight.
For instance,~\cite{SCHUTTLER2000} presents a general study and proposes some tuning rules.
Paper~\cite{COPPENS2011} compares different methods for tuning the parameters of the PK model in children.
Some other works focus on on-line identification, using data acquired during the surgical procedure.
Often, these works consider simplified PD models.
For instance,~\cite{SARTORI2006} uses a Kalman filter for the on-line identification of some of the model parameters. Also~\cite{DASILVA2012} uses the same approach for the identification of two parameters in a Single-Input-Single-Output Wiener model. Work~\cite{BIBIAN2006} uses a simplified first-order plus delay transfer function for the PK model.
In~\cite{SILVA2009}, a hybrid identification of the individual patient dynamics is employed.
Another study~\cite{HAHN2012} adopts a different model that directly correlates the propofol infusion rate and the clinical effect. In contrast, paper~\cite{LIN2004} considers piecewise linear models. Work~\cite{DASILVA2010} presents an on-line identification method based on a simplified model with four parameters, that also considers the analgesic drug.
In~\cite{DASILVA2010}, the authors point out that simple models often outperform more complex ones, due to the presence of noise, and the limited input-output data available.
Work~\cite{NOGUEIRA2013} proposes an identification procedure for the aforementioned model parameters.
Work~\cite{DASILVA2011} uses Prediction Error Method algorithms for the identification of a Multiple-Input-Single-Output system describing the action of propofol and remifentanil on the BIS signal. Paper~\cite{MERIGO2017} shows that a reduced PK model offers good prediction results, with the advantage of a lower complexity. 
Finally,~\cite{GUO2018} estimates a Wiener model parameters with an Extended Kalman filter and shows its application by testing a PID controller on a set of synthetic patients data.


\subsection{PK/PD model} 
We model the concentration and the effect of the hypnotic agent by a PK/PD model with three compartments:
\begin{equation}\label{PK/PD system}
{\small
\begin{cases}
\dot{q}_1(t) = -(k_{10}\!+\!k_{12}\!+\!k_{13})q_1(t) + k_{21}q_2(t) + k_{31}q_3(t) + v(t) \\
\dot{q}_2(t) = k_{12}q_1(t)-k_{21}q_2(t) \\
\dot{q}_3(t) = k_{13}q_1(t)-k_{31}q_3(t) \\
\dot{C_e}(t) = k_{1e}(q_1(t)/V_1)-k_{e0}C_e(t)
\end{cases}}\!\!\!\!\!\!\!\!
\end{equation}

In system~\eqref{PK/PD system}, $q_1$, $q_2$, $q_3$ are the drug masses, expressed in \unit{m\gram} in the three compartments. Namely, $q_1$ refers to the primary compartment (blood and liver), $q_2$ to the fast compartment (muscles and viscera), and $q_3$ to the slow one (fat and bones).
The input $v$ is the propofol mass-flow, expressed in \unit{m\gram\per\second}.
Variable $C_e$ is the effect-site concentration, expressed in \unit{m\gram\per\liter}. It is obtained from $q_1$ by applying a first-order low-pass filter.
The parameters of system~\eqref{PK/PD system} are the transfer rates $k_{ij}$, for $i, j \in \{1, 2, 3\}$, and the drug elimination rates $k_{10}$, $k_{e0}$, expressed in \unit{s^{-1}}. 
The measured output is the $BIS$ value.
The latter is an algebraic function of $C_e$, given by the following  \textit{Hill function}:
\begin{equation}\label{BIS}
BIS(t)=g_{\gamma,E_{\max}}(C_e(t))=E_0-E_{\textrm{max}}\left(\frac{C_e(t)^{\gamma}}{C_e(t)^{\gamma}+C_{e50}^{\gamma}}\right),
\end{equation}
where $C_{e50}$ is the effect-site concentration that corresponds to half of the maximum effect.
At each time $t$, $BIS(t)$ belongs to range $[E_0 - E_{\max}, E_0]$. Ideally, during most clinical procedures, the anesthesiologist should dose propofol to keep the BIS in range $[40,60]$.
Constant $E_0$ represents the BIS level of a fully awake and alert patient.
$E_0$ can be measured before drug infusion. Instead, $E_0 - E_{\max}$ is the $BIS$ level corresponding to a very large drug infusion. 
The higher the value, the more sensitive the patient is to the effect of propofol.

\begin{figure}
\centering
\includegraphics[width=\columnwidth,clip=true,trim=20 6 35 8]{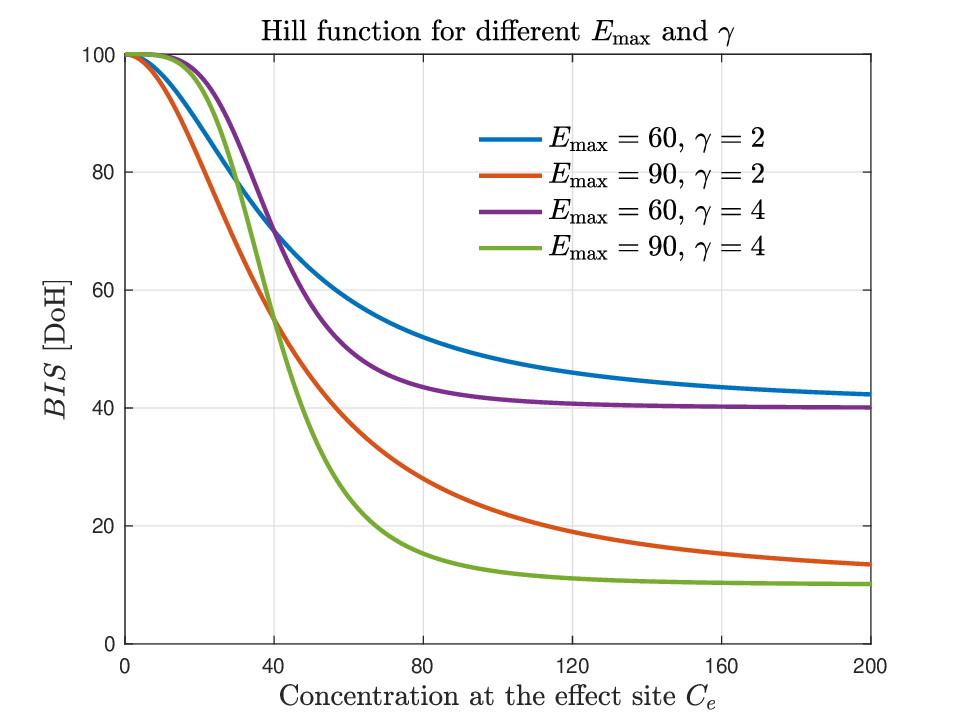}
\caption{Hill function plot for different identification parameters. Fixed constants are set as $E_0=100$ and $C_{e50}=40$.}
\label{figure:hill_function_example}
\end{figure}

\begin{figure}
\centering
\includegraphics[width=\columnwidth,clip=true,trim=25 30 35 15]{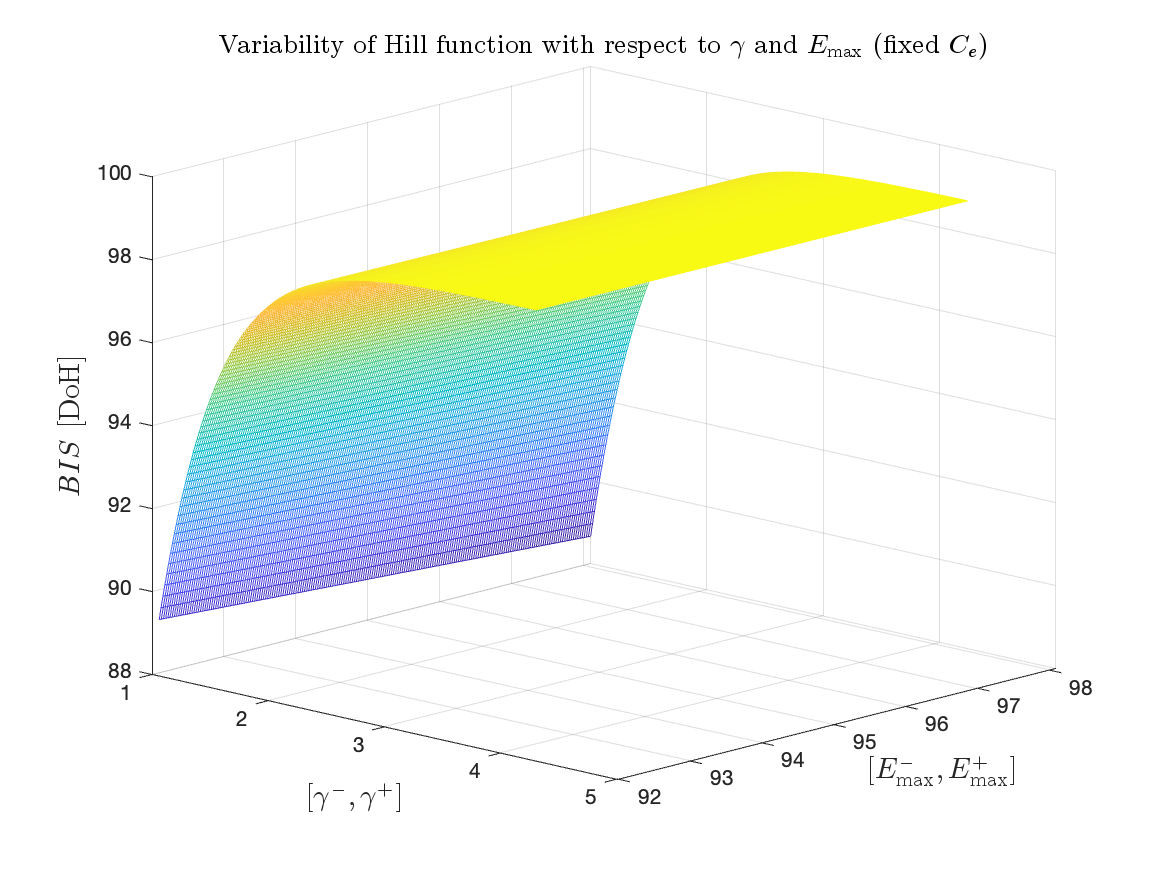}
\caption{Plot of $BIS$ for fixed $E_0$, $C_{e50}$ and $C_e$, as a function of $\gamma$ and $E_{\max}$.}
\label{figure:hill_function_parameters}
\end{figure}

The exponent $\gamma$ controls the patient's sensitivity to the hypnotic agent.
Parameter $\gamma$ can vary significantly among different patients. It is usually assumed that $\gamma>1$.
Figure~\ref{figure:hill_function_example} shows how $E_{\max}$ and $\gamma$ influence the BIS level, as a function of $C_e$.
Figure~\ref{figure:hill_function_parameters} shows, for a fixed effect site concentration $C_e$, how the $BIS$ value depends on $\gamma$ and $E_{\max}$. Note that the dependence of $BIS$ on $\gamma$ is strongly nonlinear, while the dependence on $E_{\max}$ is mostly linear.

\section{Problem formulation}
\subsection{Reinterpretation as a Wiener model}
System~\eqref{PK/PD system}--\eqref{BIS} has the structure of a Wiener model.
Indeed, it consists of the fourth-order linear system~\eqref{PK/PD system}, with input $v$ and output $C_e$, followed by Hill function~\eqref{BIS}. Let $T$ be a sampling period and set, for $k \in \mathbb{Z}$,
$u(k)=v(k T)$, $c(k)=\frac{C_e(kT)}{C_{e50}}$.
That is, we sample $u$ and $C_e$ with period $T$, and normalize $C_e$, dividing it by $C_{e50}$.
Then, the solution of linear system~\eqref{PK/PD system} satisfies a fourth-order ARX model~\cite{LJUNG1999}
\begin{equation}\label{ARX model 1}
{\small
\begin{aligned}
c(k)=	& -\alpha_1 c(k-1)-\alpha_2 c(k-2)-\alpha_3 c(k-3)-\alpha_4 c(k-4)\\
	&+\beta_1 u(k-1)+\beta_2 u(k-2)+\beta_3 u(k-3)+\beta_4 u(k-4).
\end{aligned}
}
\end{equation}
Let $y(k)=BIS(k T)$, then, we can write
\begin{equation}\label{eqn_hill_disc}
y(k)=g_{\gamma,E_{\max}} (c(k))=E_0-E_{\textrm{max}}\left(\frac{c(k)^{\gamma}}{1+ c(k)^{\gamma}}\right).
\end{equation}

\subsection{Formulation of the identification problem}
We assume that at the initial time, the drug concentration at the effect site is zero c(0) = 0, and we know u(k) and y(k) for $k \in \{0,\ldots,n\}$.
Null initial effect site concentration implies that $E_0=y(0)$, meaning that this parameter can be considered as known. Conversely, we do not know $E_{\max}$ and $\gamma$, but we can assume they belong to known, sufficiently large, intervals. That is, there exists a set $B_0=[E_{\max}^-,E_{\max}^+] \times [\gamma^-,\gamma^+]$ such that $(E_{\max}, \gamma) \in B_0$.
Also, we assume that Hill function~$g_{\gamma,E_{\max}}$ is invertible for all values of $(E_{\max},\gamma) \in B_0$. This is true if
\begin{equation}\label{assumption on emax interval}
(\forall E_{\max}\in [E_{\max}^-,E_{\max}^+])\ (\forall k\in\mathbb{Z})\ y(k)-E_{0}+E_{\max}>0.
\end{equation}
We want to identify the parameters $p_{\ell}=(\alpha_1,\ldots,\alpha_4, \beta_1,\ldots,\beta_4)$ of ARX model~\eqref{ARX model 1}, and $p=(\gamma, E_{\max})$ of Hill function~\eqref{BIS}.
Define the full set of parameters $p_f=p_{\ell}\times p$ and $B_f=\Real^{8} \times B_0$.
Then, we consider the following minimization problem
\begin{equation}\label{prob_iniz}
\begin{aligned}
\underset{p_f \in B_f}{\min}
&  \sum_{k=4}^n \left( c(k)+\sum_{i=1}^{4} \alpha_i c(k-i) -\sum_{i=1}^4 \beta_i u(k-i) \right)^2\\
\text{subject to} & \qquad c(k)=g_{\gamma,E_{\max}}^{-1} (y(k)) 
\end{aligned}
\end{equation}

Note that $g_{\gamma,E_{\max}}^{-1} (y(k))$ represents the normalized effect-site concentration that corresponds to the BIS value $y(k)$, according to parameters $\gamma$ and $E_{\max}$ of the Hill function.
The objective function of Problem~\eqref{prob_iniz} is the sum of the squared one-step ahead prediction errors, as often done in ARX identification.
Anyway, differently from standard ARX identification, function $c(k)$ is not known, but estimated by inverting the parameterized nonlinear function $g$.

We can extend Problem~\eqref{prob_iniz} to more general Wiener models, composed of an ARX model of order $(N,M)$, followed by a parameterized invertible algebraic system.
In the following, let $p_{\ell}=(\alpha_1,\ldots,\alpha_N,\beta_1,\ldots,\beta_M)$ and $p \in B_0$, $B_f=\Real^{N+M} \times B_0$, and $p_f=(p_{\ell},p)$. Consider problem
\begin{equation}
\label{prob_gen}
\begin{gathered}
\underset{p_f \in B_f}{\min}
\sum_{k=\max\{N,M\}}^n \!\!\left( c(k)+\sum_{i=1}^{N} \alpha_i c(k-i) -\sum_{i=1}^M \beta_i u(k-i) \right)^{\!\!\!2}\\
\hspace{-90pt}\text{subject to} \qquad c(k)=g_{p}^{-1} (y(k)),
\end{gathered}
\end{equation}
where we assume that $g$ be invertible for each $p \in B_0$.

\subsection{Reduction to a nonlinear regression problem}
Problem~\eqref{prob_gen} is a special case of the following nonlinear regression problem:
\begin{equation}
\label{prob_nonl_regr}
\underset{x \in \Real^n, p \in B_0}{\min} \left\| A(p) \vett{1\\x}\right\|^2 ,
\end{equation}
where $A: \mathcal{Q} \subseteq \Real^q \to \Real^{m \times {(n+1)}}$ is a $\mathcal{C}^2$ function, and $B_0 \subset \mathcal{Q}$ is a box, while $x \in \Real^n$.
If $A$ does not depend on $p$, then Problem~\eqref{prob_nonl_regr} is a standard linear regression.

To reduce Problem~\eqref{prob_gen} to form~\eqref{prob_nonl_regr}, we first substitute the nonlinear constraint in the objective function.
Set $f_p(k)=g_p^{-1} (y(k))$, define the error
\[e(k)= f_p(k)+\sum_{i=1}^{N} \alpha_i f_p(k-i) -\sum_{i=1}^M \beta_i u(k-i),\] 
and set $e=(e(\max\{N,M\}),\ldots,e(n))$.
Then, the objective function in~\eqref{prob_gen} corresponds to $\|e\|^2$.
Define  $x=\left[\alpha_1 ,\ldots ,\alpha_N ,\beta_1,\ldots,\beta_M \right]^T$.
In this way, $x$ represents the parameters of the ARX model.
The matrix in~\eqref{prob_nonl_regr} can be seen as a concatenation of two Toeplitz matrices, that is
\begin{equation}\label{eqn_def_A}
A(p)=[F(p),\;U]
\end{equation}
where, setting $\ell=\max\{M,N\}$
\[
F(p)=
\left[
\begin{matrix}
f_{p}(\ell) & f_{p}(\ell-1) & \cdots & f_{p}(\ell-N)\\
f_{p}(\ell+1) & f_{p}(\ell) & \cdots & f_{p}(\ell+1-N)\\
\vdots & \vdots & \ddots & \vdots\\
f_{p}(n) & f_{p}(n-1) & \cdots &  f_{p}(n-N) 
\end{matrix}
\right]
\]
and
\[
U=
\left[
\begin{matrix}
u(\ell-1) & \cdots & u(\ell-M)\\
u(\ell) & \cdots & u(\ell-M+1)\\
\vdots & \ddots & \vdots\\
u(n-1) & \cdots & u(n-1-M)
\end{matrix}
\right].
\]
Thanks to the previous definitions, we have that $e=A(p) \vett{1\\x}$, and Problem~\eqref{prob_gen} reduces to form~\eqref{prob_nonl_regr}.

\subsection{Statement of contribution}
The identification of the parameters of the PK/PD model is a challenging problem. Indeed, as said, many authors consider simplified models with less parameters. Often, the identification of linear models is based on the minimization of the one-step ahead prediction error. However, due the nonlinearity of the Hill function, this problem becomes non-convex for the PK/PD model. It is possible to use local search methods, but these do not guarantee finding the globally optimal model. 

With respect to existing literature, the main contribution of this work is the development of a global optimization method that guarantees finding the parameters for the PK/PD model that minimize the prediction error. In more detail:
\begin{itemize}
\item In Section~\ref{sec_bnb}, we present a Branch and Bound (BnB) method for solving a class of nonlinear regression problems, of form~\eqref{prob_nonl_regr}.
In particular, our algorithm exploits an efficient relaxation of this problem.
\item We apply the proposed method to the identification of a class of Wiener models, including the PK/PD model of hypnotic agents in general anesthesia.
\end{itemize}



\section{A BnB method for solving Problem~\eqref{prob_nonl_regr}}\label{sec_bnb}
In general, due to dependence of $A(p)$ on $p$, Problem~\eqref{prob_nonl_regr} is nonlinear and non-convex.
In this section we propose a BnB approach for its solution.

Let $\mathcal{B}$ be the set of boxes included in $B_0$. Define function 
$f^*:\mathcal{B} \to \Real$ as
\begin{equation}
\label{eq:fstar}
f^*(B)= \underset{x \in \Real^n, p \in B}{\text{min}} \hat f(p,x)=\left\| A(p) \vett{1\\x}\right\|^2 .
\end{equation}
Further, set $f(p)=\underset{x \in \Real^n}{\text{min}} \hat f(p,x)$.
Assume that there exists a function $L:\mathcal{B} \to \Real$, such
that,
\begin{equation}\label{eqn_def_lb}
(\forall B \in \mathcal{B})\ L(B) \leq f^*(B).
\end{equation}
We will call any $L$ satisfying~\eqref{eqn_def_lb} a
\emph{lower bound} function of $f^*$.
Further, let function $r:\mathcal{B} \to \Real^q$ be such that $(\forall B \in \mathcal{B})\ r(B) \in B$.  
Function $r$ returns a point within box $B$ (in our
numerical experiments we always return the center of the box).
The optimal solution of Problem~\eqref{prob_nonl_regr} can be found
with the standard BnB Algorithm~\ref{alg_bnb} adapted
from~\cite[p.~18]{BrAndBound}. The algorithm uses a binary
tree whose nodes are associated to a restriction of
Problem~\eqref{prob_nonl_regr} to a box, obtained by
recursively splitting the initial box $B_0$. Input parameter $\epsilon$
represents the maximum relative allowed error on the objective function for
the optimal solution, and the output variable $x^*$ is an
approximation of the optimal solution with relative tolerance $\epsilon$.
In Algorithm~\ref{alg_bnb}, function $\delta: \mathcal{B} \to \Real$ is
used to define the exploration policy for set $\zeta$. For instance,
in a best first search strategy, the node with the lowest lower bound
is the next to be processed, so that $\delta(\eta)=L(\eta)$
(this is also the choice that we made throughout the paper).
\begin{algorithm}[!b]
\caption{Main BnB algorithm}
\label{alg_bnb}
\begin{algorithmic}
\State Input: 
\State $\epsilon$: solution tolerance
\State Output: $x^*$: optimal solution 
\begin{enumerate}
\State Let $\zeta$ be a list of boxes and initialize $\zeta=\{B_0\}$.
\State Set $UB=f(r(B_0))$, and $x^*=r(B_0)$.
\State If $\zeta=\varnothing$, stop. Else set $\delta_{\min} =
  \min \{ \delta(\eta)\ |\ \eta \in \zeta \}$.
\State Select a box $\eta \in \zeta$, with $\delta(\eta)=\delta_{\min}$
  and split it into two equal smaller sub-boxes $\eta_{1}$, $\eta_{2}$ along
  the dimension of maximum length. \label{subd}
\State Delete $\eta$ from $\zeta$ and add $\eta_{1}$ and $\eta_{2}$ to $\zeta$.
\State Update $UB=\min\{UB, f(r(\eta_{1})), f(r(\eta_{2}))\}$.
If $UB= f(r(\eta_j))$ with $j \in \{1,2\}$, set
$x^*= r(\eta_{j})$.
\State Let $\zeta=\zeta \setminus \{\kappa \in \zeta \mid UB\leq (1+\epsilon) L(\kappa)\}.$
\State Return to Step 3.
\end{enumerate}
\end{algorithmic}
\end{algorithm}
Note that the choice of the lower bound function
$L$ is critical to
efficiency of Algorithm~\ref{alg_bnb}.
The following property on $L$ guarantees that
Algorithm~\ref{alg_bnb} converges to a solution of
Problem~\eqref{prob_nonl_regr}, with relative tolerance $\epsilon$.
\begin{equation}
\label{eqn_cond_on_lb}
\lim_{\sigma(B) \to 0}  \left(L(B)-f^*(B) \right)=0\,,
\end{equation}
where $\sigma(B)$ denotes the diameter of box $B$ (note that the subdivision rule employed at line~\ref{subd} of Algorithm~\ref{alg_bnb} guarantees that $\sigma(B) \to 0$ if the stopping rule of the algorithm is removed).

We a propose a lower bound for $f^*(B)$ in~\eqref{eq:fstar}.
Given $\bar p \in B$, we rewrite objective function~\eqref{eq:fstar} as
\[\begin{aligned}
  f(p,x) = & \left \|\left(A(p)-A(\bar p)+ A(\bar p)\right)\vett{1\\x}\right \|^2 = \\
  = & \left \|A(\bar p) \vett{1\\x}\right \|^2 + \left \|\left(A(p)- A(\bar p)\right)  \vett{1\\x}\right \|^2 + \\
 & + 2 \vett{1, x^T} \left(A(p)-A(\bar p)\right)^T A(\bar p) \vett{1\\x}.
\end{aligned}
\]

Hence, the next problem gives a lower bound for $f^*(B)$
\begin{equation}
\label{prob_nonl_regr_box_lb}
\underset{\substack{x \in \Real^n\\ p \in B}}{\text{min}}
\left \|A(\bar p) \vett{1\\x}\right \|^2
 + 2 \vett{1, x^T} \left(A(p)-A(\bar p)\right)^T A(\bar p) \vett{1\\x}.
\end{equation}

Define $\O(p)=A(p)- A(\bar p) - \nabla A(\bar p) (p-\bar p)$.
Note that $\O(p)$ is the remainder of the first order Taylor expansion of $A(p)$ at $\bar p$.
Then,
\[
\begin{gathered}
\vett{1,x^T} \left(A(p)-A(\bar p)\right)^T A(\bar p) \vett{1\\x}=\\
= \vett{1,x^T} (p-\bar p) \nabla A(\bar p)^T A(\bar p) \vett{1\\x}+ \vett{1,x^T} \O(p)^T A(\bar p) \vett{1\\x}.
\end{gathered}
\]

To find a bound on $\vett{1,x^T} \O(p)^T A(\bar p) \vett{1\\x}$ we use the following property.

\begin{proposition}\label{prop_bound}
  Let $M,N \in \Real^{m \times n}$, let $k \in \Real$, with $k>0$, then
  \[
M^T N+ N^T M \geq -\frac{1}{k}N^T N - k M^T M.
  \]
\end{proposition}
\begin{proof}
  Note that
  \[
\left(\frac{N}{k}+M\right)^{\!T}\!\!\!\! \left(\frac{N}{k}+M\right)\geq 0,
\]
then
\[
\frac{1}{k^2}N^T N + \frac{1}{k} \left(N^T M + M^T N\right) + M^T M \geq 0,
\]
and
\[
N^T M + M^T N \geq - \frac{1}{k} N^T N - k M^T M.
\]\vspace{-22pt}

\end{proof}

Let $r_B$ ve such that $r_B \geq \max_{p \in B} \|\mathcal{O}(p)\|$. 
We apply Proposition~\ref{prop_bound} with $N=A(\bar p)$, $M=\O(p)$.
Then, for all $p \in B$, $k>0$
\begin{equation}
  \label{eqn_bound_termine}
  \begin{gathered}
  \O(p)^T A(\bar p) + A(\bar p)^T \O(p) \geq \\
  \geq-\frac{1}{k} A(\bar p)^TA(\bar p)-k \O(p)^T \O(p) \\
  \geq  M_{\bar p,B,k} \geq - \frac{1}{k} A(\bar p)^TA(\bar p)- k r_B^2.
  \end{gathered}
\end{equation}

Note that bound~\eqref{eqn_bound_termine} holds for any $k>0$.

We can find $r_B$ using the following property.

\begin{proposition}
For $i \in \{1,\ldots,m\}$, $j \in \{1,\ldots,n\}$, let $H_{i,j}: B \to \Real^{q \times q}$ be
the Hessian matrix of $A_{i,j}$ (the element of $A$ at row $i$ and column $j$) and assume that there exists a constant $R_{i,j}$ such that, for all $p \in B$
\begin{equation}\label{upper bound for hessian}
\|H_{i,j}(p)\| \leq R_{i,j},
\end{equation}
then, for all $p \in B$
\[
\|\mathcal{O}(p)\|^2 \leq \frac{1}{4}\sum_{i \in \{1,\ldots,m\}, j \in \{1,\ldots,n\}} R_{i,j}^2 d(\bar p, B)^4,
\]
where $d(\bar p,B)$ is the maximum distance of $\bar p$ to set $B$, that is
\[
d(\bar p, B)=\max_{p \in B} \|p-\bar p\|.
\]
\end{proposition}

\begin{proof}
  For any $i,j$, from the formula for the Lagrange remainder, there exists 
  $\hat p\in [p,\bar{p}]\subset B$ such that $\mathcal{O}_{ij}(p)=\frac{1}{2}(p-\bar p)^T H_{i,j} (\hat p) (p-\bar p)$. Hence
  $|\mathcal{O}_{ij}(p)| \leq \frac{1}{2} d(\bar p,B)^2 R_{i,j}$.
  The thesis follows by bounding the $2$-norm of $\mathcal{O}(p)$ by its Frobenius norm.
\end{proof}


Then, the following is a lower bound for~\eqref{prob_nonl_regr_box_lb}, and, hence, for $f^*(B)$
\begin{equation}\label{prob_nonl_regr_box_lb_rel}
\begin{aligned}
L(B)=& \underset{x \in \Real^n, p \in B}{\text{min}}
\vett{1,x^T} (A^T(\bar p) A(\bar p)+M_{\bar p,B,k}) \vett{1\\x} + \\
& + 2 \vett{1,x^T} (\nabla A(\bar p) (p-\bar p))^T A(\bar p) \vett{1\\x},
\end{aligned}
\end{equation}
Problem~\eqref{prob_nonl_regr_box_lb_rel} is linear with respect to $p$.
Hence, the minimum with respect to $p$ is attained at a vertex of box $B$. Let $V$ be the set of vertices of $B$, and define function
\begin{equation}\label{prob_nonl_regr_box_lb_rel_p}
\begin{aligned}
L_{p,k}(B)=& \underset{x \in \Real^n}{\text{min}}
\vett{1,x^T} (A^T(\bar p) A(\bar p)+M_{\bar p,B,k}) \vett{1\\x} + \\
&+ 2 \vett{1,x^T} (\nabla A(\bar p) (p-\bar p))^T A(\bar p) \vett{1\\x}.
\end{aligned}
\end{equation}

Then, $L(B)=\min_{p \in V} L_{p}(B)$.
The computation of $L_p(B)$ is a direct consequence of the following algebraic decomposition for bilinear forms:
\begin{proposition}\label{algebraic decomposition}
Let $m,n$ be positive integers and $A,B\in \Real^{m \times (n+1)}$. It is possible to find $Q\in \Real^{n \times n}$, $c\in\Real^{n}$ and $d\in \Real$ such that for all $x\in\Real^{n}$ it holds
\begin{equation}\label{product split}
\vett{1,x^T}A^T B\vett{1\\x}=x^T Q x + c^Tx +d
\end{equation}
\end{proposition}
\begin{proof}
Decompose $A$, $B$ as
  \[
A=\vett{
a_{11} &A_{12}\\
A_{21} & A_{22}\\
},\quad
B=\vett{
b_{11} &B_{12}\\
B_{21} & B_{22}\\
},
\]
where the first diagonal elements $a_{11}$ and $b_{11}$ are highlighted, then it holds
\begin{align*}
Q & = A_{12}^T B_{12}+A_{22} B_{22},\\
c & = a_{11}B_{12}^T+B_{22}^TA_{21}+b_{11}A_{12}^T+ A_{22}B_{21},\\
d & = a_{11}b_{11}+A_{22}^TB_{21}.
\end{align*}\vspace{-28pt}

\end{proof}
Thanks to Proposition~\ref{algebraic decomposition}, and observing that for a fixed $p\in V$ objective function~\eqref{prob_nonl_regr_box_lb_rel_p} is a sum of bilinear forms as in the left hand side of~\eqref{product split}, we are able to rewrite~\eqref{prob_nonl_regr_box_lb_rel_p} in the following equivalent form
$$L_{p,k}(B)=\underset{x \in \Real^n}{\text{min}}\: x^T Q_{p,B,k} x + c_{p,B,k} x +d_{p,B,k}.$$
By construction, $Q_{p,B,k}$ is symmetric: as a consequence, the Hessian matrix of the objective function of the above minimization problem is $2 Q_{p,B,k} $. Hence, if $Q_{p,B,k}$ is indefinite, then $L_{p,B,k}(B)=-\infty$ (and the computed lower bound is useless). 
The same holds true if $Q_{p,B,k}$ is semidefinite positive and $c_{p,B,k} x$ is not null over the null space of $Q_{p,B,k}$. In general, we will set $L_{p,B,k}(B)=-\infty$ when $Q_{p,B,k}$ is not positive definite. 
Otherwise, if $Q_{p,B,k}$ is positive definite, $L_{p,B,k}(B)$ is the optimal value of a strictly convex quadratic problem and is computable in closed form:
$$L_{p,k}(B)=x^{*T} Q_{p,B,k} x^* + c_{p,B,k} x^* +d_{p,B,k},$$
with $x^*$ solution of $Q_{p,B,k}x^*=-\frac{c_{p,B,k}}{2}$.

Note that lower bound $L_{p,k}(B)$ depends on $k>0$. Hence, we compute the best lower bound by maximizing $L_{p,k}(B)$ with respect to $k$.


\section{Application to the identification of the Wiener model}
As said,~\eqref{prob_gen} is a nonlinear regression problem characterized by the structure defined in~\eqref{prob_nonl_regr} and by the matrix $A$ defined in~\eqref{eqn_def_A}. To compute bound~\eqref{prob_nonl_regr_box_lb_rel}, we need to find an upper bound for $\|H_{i,j}(p)\|$, the Hessian of the element of $A(p)$ at row $i$ and column $j$. Note that in $A(p)$, defined in~\eqref{eqn_def_A}, only Toeplitz block $F(p)$ depends on $p$.  Setting $k=i+j-1$ and $a_k=(E0-y(k))/(E_{\max}-E0+y(k))$, $(\forall i \in \{1,\ldots,T\})\ (j \in \{1,\ldots,N+1\})$, the elements of $F(p)$ are
\begin{equation}\label{inverse of hill function}
A_{i,j}(p)=f_p(k)=a_k^{\frac{1}{\gamma}},
\end{equation}
resulting from the inversion of the Hill function~\eqref{eqn_hill_disc} at the $k$-th sample instant. We highlight the matrix elements just for the subsets of $N+1$ column indexes associated to the block $F(p)$, since the entries of the block $U$ are independent on the identification parameters and therefore their Hessians result trivial:
\begin{equation}\label{explicit hessian}
H_{i,j}(p)=\left[
\begin{matrix}
\partial_\gamma^2 (f_p(k)) & \partial_{E_{\max}}\partial_{\gamma} (f_p(k))\\
\partial_{E_{\max}}\partial_{\gamma} (f_p(k)) & \partial_{E_{\max}}^2 (f_p(k))\\
\end{matrix}
\right]
\end{equation}
with
\begin{align*}
&\partial_\gamma^2 (f_p(k))=\log(a_k)a_k^{\frac{1}{\gamma}}\frac{(2\gamma+\log(a_k))}{\gamma^4},\\
&\partial_{E_{\max}}\partial_{\gamma}(f_p(k))=a_k^{\frac{1}{\gamma}}\frac{(\gamma+\log(a_k))}{\gamma^3(y(k)-E0+E_{\max})},\\
&\partial_{E_{\max}}^2 (f_p(k))=a_k^{\frac{1}{\gamma}}\frac{(\gamma+1)}{\gamma^2(y(k)-E0+E_{\max})^2}.
\end{align*}
As mentioned previously, we can compute as suitable upper bound $R_{i,j}$ in~\eqref{upper bound for hessian} starting from the Frobenius norm of~\eqref{explicit hessian}:
\begin{align*}
\Vert H_{i,j}(p)\Vert_{\mathcal{F}}&=\sqrt{\sum_{p_1,p_2=\gamma,E_{\max}}(\partial_{p_1}\partial_{p_2}f_p(k))^2}\\
&\leq \sqrt{\sum_{p_1,p_2=\gamma,E_{\max}}\max_{p\in B}(\partial_{p_1}\partial_{p_2}f_p(k))^2}=R_{i,j}.
\end{align*}
During the identification we always assume $B$ as a compact domain contained in the primary identification interval $B_0$, constructed imposing condition~\eqref{assumption on emax interval}. This assumption and the fact that we always search for an exponent $\gamma>1$, guarantee continuity for $\partial_{p_1}\partial_{p_2}f_p(k)$ in $B$. Therefore, the Hessian entries have a maximum that can be explicitly computed with a further study of the gradient of these functions, which we omit in this work for sake of simplicity.

\section{Experimental results}

\subsection{Patients database}
We considered a standard patients database of $12$ individuals, differentiated by age, height, weight and gender (see~\cite{IONESCU2008}).
We added a thirteenth patient, determined as the algebraic average of the other individuals. Table~\ref{table:data_patients} presents the patients parameters. Note that their variability is quite large.
We computed the parameters of the PK/PD model~\eqref{PK/PD system} with the method in~\cite{SCHNIDER1998}.

 We assumed that $\gamma \in [1,8]$ and $E_{\max} \in [40,160]$. This corresponds to the initial box $B_0=[1,8]\times[40,160]$.


\begin{table}
\center{
\begin{tabular}{|c|c|c|c|c|c|c|c|c|}
\hline
  \!id\! & \!age\! & \!\!height\!\! & \!\!weight\!\! & \!\!gender\!\! & $C_{e50}$ & $\gamma_{\textrm{ob}}$ & $E_0$ & $E_{\max,\textrm{ob}}$  \\
  \hline
  1 &  40 & 163 & 54 & f & 6.33 & 2.24 & 98.8 & 94.10\\
  2 &  36 & 163 & 50 & f & 6.76 & 4.29 & 98.6 & 86.00\\
  3 &  28 & 164 & 52 & f & 8.44 & 4.10 & 91.2 & 80.70\\
  4 &  50 & 163 & 83 & f & 6.44 & 2.18 & 95.9 & 102.00\\
  5 &  28 & 164 & 60 & m & 4.93 & 2.46 & 94.7 & 85.30\\
  6 &  43 & 163 & 59 & f & 12.00& 2.42 & 90.2 & 147.00\\
  7 &  37 & 187 & 75 & m & 8.02 & 2.10 & 92.0 & 104.00\\
  8 &  38 & 174 & 80 & f & 6.56 & 4.12 & 95.5 & 76.40\\
  9 &  41 & 170 & 70 & f & 6.15 & 6.89 & 89.2 & 63.80\\
  10&  37 & 167 & 58 & f &13.70 & 1.65 & 83.1 & 151.00\\
  11&  42 & 179 & 78 & m & 4.82 & 1.85 & 91.8 & 77.90\\
  12&  34 & 172 & 58 & f & 4.95 & 1.84 & 96.2 & 90.80\\
  13&  38 & 169 & 65 & f & 7.42 & 3.00 & 93.1 & 96.58\\
  \hline
\end{tabular}}
\caption{Patients' data.}
\label{table:data_patients}
\end{table}

\subsection{Numerical tests}
We implemented the BnB algorithm~\ref{alg_bnb} in Matlab.
We consider and interval of induction of $300$ second with sample period $T=1$s and considered the following input
We used a piecewise constant input
\[
v(t)=
\begin{cases}
10, 0\leq t < 10 \\
3, 0 \leq t < 25 \\
0, t \geq 25
\end{cases}.
\]
representing a bolus of propofol administrated in the first $10$ seconds, followed by a period of $15$ seconds of lower infusion.
The choice of input $v$ is critical to the identification process. It is difficult to find an input suitable for all patients in Table~\ref{table:data_patients}, due to the large parameter variability.

 We set the order of the  ARX model~\eqref{prob_gen} to $N=M$, and we considered $N \in \{2,3\}$. Note that we did not consider the full order $N=M=4$, since the input signal is too short to have a significant contribution of the dynamics of the slow component.

Solving~\eqref{PK/PD system} and using Hill function~\eqref{BIS}, we computed the BIS sampled signal concentrations $y_{id}(k)$, where $id \in \{1,\ldots,12\}$ is the patient number.
 Table~\ref{table:error_object_function} presents the results of the numerical experiments. In particular, the first column is the patient's Id, the second and third are the order of the ARX model, the fourth column is the minimum of objective function~\eqref{prob_nonl_regr}. The fifth columns is the total number of computed lower bounds, and the last column is the norm of the difference between the estimated value $\hat p$ of the parameters of the Hill function (that is, $\gamma$ and $E_{\max}$) and their true values $p^*$. We committed a larger error on patient number $9$. This is probably due to the fact that this patient has very peculiar parameters ($\gamma=6.89$, $E_{\max}=63.80$). Near these values, the sensitivity of the BIS signal to variations of these two parameters is quite small.

\begin{figure}
\centering
\includegraphics[width=\columnwidth,clip=true,trim=29 13 0 5]{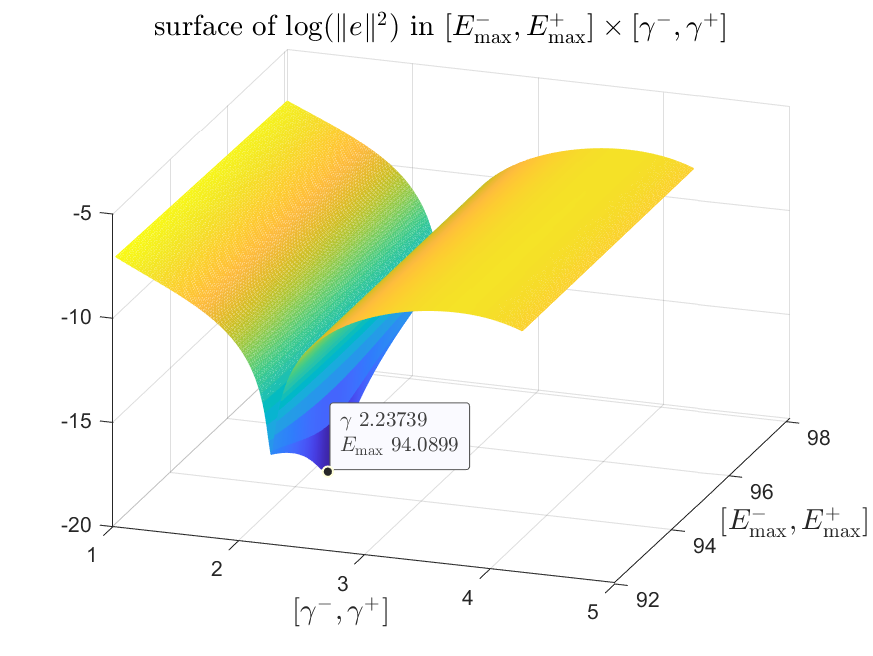}
\caption{Plot of function $h$ defined in~\eqref{eqn_for_plot}, for $M=N=2$ and Partient $Id=1$.}
\label{figure:error_surface}
\end{figure}

 Figure~\ref{figure:error_surface} shows a plot of function

 \begin{equation}
   \label{eqn_for_plot}
h(p) = \underset{x \in \Real^n}{\min} \log \left\| A(p) \vett{1\\x}\right\|^2 .
\end{equation}

That is, $h(p)$ is the logarithm of the minimum error resulting from the solution of problem~\eqref{prob_nonl_regr} with fixed $p$ (that can be solved by linear regression). For this plot, we chose $M,N=2$ and considered the first patient ($Id=1$). The minimum is reached approximatively at the optimal values $\gamma=2.24$ and $E_{\max}=94.1$.

Note that an ARX model of order $(2,2)$ is sufficient for the correct identification of the Hill function parameters. This is probably related to the fact that identification is based on a short signal ($300$ seconds), and the drug concentration at the effect site depends mainly on the kinetics of the primary compartment. The kinetics of fast and slow compartments are almost irrelevant in this short time scale. This is in accordance with existing literature. Indeed, as mentioned in the Introduction, various authors showed that, in many cases, a system of order two is sufficient for a good approximation the PK model.

Table~\ref{table:error_object_function} collects data experiments for the evaluation of the minimum of the object function in~\eqref{prob_nonl_regr}. As the reader can deduce, globally for all the tested patients, with four states the error is numerically near to be null, indicating an exact identification of the nonlinear parameters studied, however, at the expense of a greater number $n_s$ of subsets of the initial identification box explored in the branching phase. 

\begin{table}
\center{
  \begin{tabular}{|c|c|c|r|r|l|}
    \hline
Patient Id & $N$ & $M$ &  $\min \|e\|^2$ & \# LBs & $\|\hat p - p^*\|$ \\ 
\hline 
1 & 2 & 2 & $9.4384\cdot10^{-8}$ & 49635 & 0.0072816 \\ 
1 & 3 & 3 & $-3.1287\cdot10^{-10}$ & 96415 & 0.0024104 \\ 
2 & 2 & 2 & $1.4623\cdot10^{-6}$ & 39571 & 0.008462 \\ 
2 & 3 & 3 & $1.0445\cdot10^{-7}$ & 72179 & 0.066098 \\ 
3 & 2 & 2 & $1.2895\cdot10^{-6}$ & 33905 & 0.0088561 \\ 
3 & 3 & 3 & $7.8096\cdot10^{-8}$ & 65609 & 0.061555 \\ 
4 & 2 & 2 & $1.4297\cdot10^{-7}$ & 43383 & 0.0085839 \\ 
4 & 3 & 3 & $-2.6193\cdot10^{-10}$ & 91699 & 0.0018231 \\ 
5 & 2 & 2 & $1.509\cdot10^{-7}$ & 52887 & 0.0038887 \\ 
5 & 3 & 3 & $-7.4579\cdot10^{-10}$ & 94013 & 0.00062041 \\ 
6 & 2 & 2 & $1.0526\cdot10^{-7}$ & 33885 & 0.034773 \\ 
6 & 3 & 3 & $-1.2005\cdot10^{-10}$ & 77747 & 0.0043385 \\ 
7 & 2 & 2 & $1.2996\cdot10^{-7}$ & 39643 & 0.014525 \\ 
7 & 3 & 3 & $-5.748\cdot10^{-10}$ & 91511 & 0.0052605 \\ 
8 & 2 & 2 & $8.0302\cdot10^{-7}$ & 36735 & 0.0045107 \\ 
8 & 3 & 3 & $5.776\cdot10^{-8}$ & 68829 & 0.049279 \\ 
9 & 2 & 2 & $3.3758\cdot10^{-4}$ & 51339 & 0.11687 \\ 
9 & 3 & 3 & $9.9934\cdot10^{-6}$ & 42135 & 0.3826 \\ 
10 & 2 & 2 & $1.734\cdot10^{-7}$ & 31799 & 0.029549 \\ 
10 & 3 & 3 & $-5.1659\cdot10^{-10}$ & 91869 & 0.0066823 \\ 
11 & 2 & 2 & $1.0198\cdot10^{-7}$ & 52805 & 0.0052619 \\ 
11 & 3 & 3 & $-3.9654\cdot10^{-10}$ & 101757 & 0.00052213 \\ 
12 & 2 & 2 & $1.2325\cdot10^{-7}$ & 50773 & 0.0051633 \\ 
12 & 3 & 3 & $-7.2032\cdot10^{-10}$ & 102217 & 0.0005642 \\ 
13 & 2 & 2 & $1.3616\cdot10^{-7}$ & 40525 & 0.01154 \\ 
13 & 3 & 3 & $1.9645\cdot10^{-10}$ & 80423 & 0.0032022 \\ 
\hline 
  \end{tabular}}
\caption{Numerical results.}
\label{table:error_object_function}
\end{table}

\section{Conclusions and future works}

In this work we introduced a global optimization method for the identification of PK/PD model parameters. This approach ensures the minimization of the error in a non-convex setting, which is critical for accurately predicting the effects of hypnotic drugs during TIVA. In the proposed BnB method, we introduced a lower bound for the objective function which allows cutting the exploration of large portions of the parameters' domain. This method overcomes the limitations of local search methods which cannot guarantee globally optimal solutions.

By providing an accurate and precise estimate of model parameters, our approach allows anesthesiologists to tailor anesthesia procedures to individual patients more effectively. This not only reduces the risks associated with under or over-dosing hypnotic drugs, such as patient awareness or hypotension, but also improves postoperative outcomes.

In future works we plan to explore the application of the proposed optimization method to other drugs or combinations of them, and to other medical scenarios where PK/PD models are utilized (e.g., intensive care unit). Moreover, further validation of our method through clinical trials would be fundamental in assessing its effectiveness and reliability in real-life scenarios.


\addtolength{\textheight}{-3cm}   



\bibliographystyle{abbrv}
\bibliography{biblio}

\end{document}